\documentstyle[preprint,tighten,aps,floats,psfig,epsfig]{revtex}

\def\ub{{\overline{u}}}
\def\vb{{\overline{v}}}

\begin{document}

\title 
{Chiral critical behavior in two dimensions from five-loop 
renormalization-group expansions} 

\author{
P. Calabrese${}^{1}$, 
E. V. Orlov${}^2$,
P. Parruccini${}^3$, and 
A. I. Sokolov${}^2$, 
}

\address{$^1$Scuola Normale Superiore and INFN,
Piazza dei Cavalieri 7, I-56126 Pisa, Italy.}

\address{$^2$Department of Physical Electronics, Saint Petersburg 
Electrotechnical University, Professor Popov Street 5, 
St. Petersburg 197376, Russia.}

\address{$^3$Dipartimento di Fisica dell' Universit\`a di 
Pisa and INFN, Via Buonarroti 2, I-56100 Pisa, Italy. \\
{\bf e-mail: \rm 
{\tt calabres@df.unipi.it},
{\tt orlov@mail.lanck.net},
{\tt parrucci@df.unipi.it},
{\tt ais@sokol.usr.etu.spb.ru}.
}}

\date{\today}

\maketitle

\begin{abstract}
We analyze the critical behavior of two-dimensional $N$-vector spin 
systems with noncollinear order within the five-loop renormalization-group  
approximation. The structure of the RG flow is studied for different $N$ 
leading to the conclusion that the chiral fixed point governing the 
critical behavior of physical systems with $N = 2$ and $N = 3$ does not 
coincide with that given by the $1/N$ expansion. We show that the stable 
chiral fixed point for $N \le N^*$, including $N = 2$ and $N = 3$, 
turns out to be a focus. 
We give a complete characterization of the critical behavior controlled 
by this fixed point, also evaluating the subleading crossover exponents. 
The spiral-like approach of the chiral fixed point is argued to give 
rise to unusual crossover and near-critical regimes that may imitate 
varying critical exponents seen in numerous physical and computer 
experiments.
\end{abstract}


\section{Introduction}

The critical behavior of two-dimensional frustrated spin models with 
noncollinear or canted order has been the object of intensive theoretical 
and experimental studies, being of prime interest for those investigating 
layered magnetic systems with special structures and superconducting 
Josephson-junction arrays in an external magnetic field.

In physical magnets, frustration (leading to noncollinear order) may arise 
either because of the special geometry of the lattice or from the 
competition of different kinds of interactions. An example of the first 
kind is the two-dimensional triangular antiferromagnet, whereas for the 
latter case typical models are the fully frustrated $XY$ model on a 
square lattice and models with different nearest-neighbor and 
next-nearest-neighbor interactions. In these systems the Hamiltonian is 
minimized by the noncollinear configurations with a 120$^o$ spin structure. 
As a consequence, at criticality, there is a breakdown of O($N$) symmetry 
in the high-temperature phase to O($N-2$) symmetry in the ordered phase.

Field-theoretical (FT) studies of systems with noncollinear order are 
based on the O($N$)$\times$O($M$) symmetric Landau-Ginzburg-Wilson (LGW)
 Hamiltonian 
~\cite{Kawamura-88,Kawamura-98,rev-01} 
\begin{eqnarray}
{\cal H} = \int d^d x 
&& \left\{ {1\over2}
      \sum_{a} \left[ (\partial_\mu \phi_{a})^2 + r \phi_{a}^2 \right] 
+ {1\over 4!}u_0 \left( \sum_a \phi_a^2\right)^2 \right. 
 \left. + {1\over 4!}  v_0 
\sum_{a,b} \left[ ( \phi_a \cdot \phi_b)^2 - \phi_a^2\phi_b^2\right]
             \right\}, 
\label{LGWH}
\end{eqnarray}
where $\phi_a$ ($1\leq a\leq M$) are $M$ sets of $N$-component vectors. 
We will consider the case $M=2$, that, for $v_0>0$, 
describes frustrated systems with noncollinear ordering. 
Negative values of $v_0$ correspond to simple ferromagnetic or 
antiferromagnetic ordering, and to magnets with sinusoidal 
spin structures~\cite{Kawamura-98}.

The physically relevant cases are described by the Hamiltonian (\ref{LGWH}) 
with $N=2$ (frustrated $XY$ model) and $N=3$ (frustrated Heisenberg model).
Despite the very intensive theoretical and experimental efforts to 
fully understand the critical behavior of these models, the situation is  
still controversial. For the frustrated $XY$ model the strongly debated 
point is either the single critical temperature $T_c$ exists at which 
both the $SO(2)$ and the $Z_2$ symmetries are simultaneously broken or 
there are two successive phase transitions at different critical 
temperatures. In the latter case, the order at which two phase 
transitions occur and the numerical values of the critical exponents are 
doubtful too. Obviously, studying the Hamiltonian (\ref{LGWH}), we cannot 
clarify the point just mentioned, since this Hamiltonian describes only the 
chiral phase transition. In more detail, these and other relevant 
issues are reviewed in Ref. \cite{pp-01}, where one can find a 
complete list of references; we mention here only the very recent 
studies \cite{lsz-98,mclmtmm-00,clj-01,feknr-02} 
(not quoted in Ref. \cite{pp-01}).

The critical behavior of the frustrated Heisenberg antiferromagnet in two 
dimensions is clearer: it undergoes one phase transition mediated by $Z_2$ 
topological defects. Again, a complete list of references can be found 
in Ref. \cite{pp-01}. Here the debated point concerns whether the LGW 
Hamiltonian is able to keep the non trivial topological excitations 
present in this model, to the contrary of other approaches like the 
nonlinear $\sigma$ model ($NL\sigma$). In Ref. \cite{pp-01} it is claimed that it 
may be the case and we will assume it in the following. 

The LGW Hamiltonian (\ref{LGWH}) has been extensively studied in the 
framework of $\epsilon = 4 - d$ \cite{Kawamura-98,asv-95,prv-01b}, and 
by means of the $1/N$ \cite{Kawamura-98,prv-01b} expansion. The existence 
and the stability properties of the fixed points were found to depend on 
$N$ and on the spatial dimensionality \cite{Kawamura-98,rev-01}. Within 
the $\epsilon$ expansion, for sufficiently large values of $N$ the 
renormalization group equations possess four fixed points: the Gaussian 
fixed point ($v=u=0$), the $O(2N)$ ($v=0$) one and two anisotropic 
fixed points 
located in the region $u,v>0$ and usually named chiral and 
antichiral. The chiral fixed point is the only stable one.
 There is a critical dimensionality $N_c$ where the chiral 
and antichiral fixed points coalesce and disappear for $N<N_c$. In the 
last case, under the absence of any stable fixed point the system 
undergoes a weak first-order phase transition, since the associated RG 
flows run away from the region of stability of the fourth-order form in 
the free energy expansion. The three-loop estimate of $N_c$ is 
\cite{asv-95}
\begin{equation} 
N_c= 21.8-23.4 \epsilon +7.1 \epsilon^2+ O(\epsilon^3)\, ,
\end{equation}
that, after an appropriate resummation, results in $N_c > 3$ in 
three dimensions. This inequality leads to the conclusion that 
for the physical models with $N=2, 3$ the three-dimensional chiral 
transition is first order, as corroborated also by some other RG studies 
\cite{as-94,tdm-00}. To the contrary, both for $N=2$ and $N=3$ the 
highest-order six-loop calculations in three dimensions \cite{prv-01} 
reveal a strong evidence for a stable chiral fixed point that, however, 
is not related to its counterpart found within the $\epsilon$ expansion. 
The four-loop two-dimensional analysis shows an equivalent topology of 
the fixed points location in the renormalized coupling constants plane 
\cite{pp-01}. Furthermore, in a recent work \cite{CPS-02} we claimed 
that both in two and three dimensions the controversial situation may 
reflect the quite unusual mode of critical behavior of the $N$-vector 
chiral model under the physical values of $N$. It was 
shown that this critical behavior is governed by a stable fixed point 
which is a focus, attracting RG trajectories in a spiral-like manner. 
Approaching the fixed point in a nonmonotonic way, looking somewhat 
irregular, may result in a large variety of the crossover and 
near-critical regimes.

In this paper, we report the perturbative five-loop two-dimensional RG 
expansions and carry out their analysis. Some of the results for 
$N = 2$ and $N = 3$ was already anticipated in \cite{CPS-02}. We point 
out the idea that the critical behavior for large $N$ is the one 
predicted by the $\epsilon$ expansion. With decreasing $N$, the stable 
chiral fixed point becomes focuslike at some marginal value of $N$, 
$N^*$ that differs, in principle, from $N_c$ where both chiral and 
antichiral fixed points disappear. With the precision of our calculations, 
we are not able to fully clarify if there is a region of  
values of $N$, where the chiral transition is first order. 
It is worthy to note that the parallel work on the three-dimensional 
models \cite{CPS-02b} reveals a full analogy between chiral critical 
behaviors in two and three dimensions. 

The paper is organized as follows. In Sec. II we derive the perturbative 
series for the renormalization-group functions up to five-loop order. 
The resummations methods and the results of the analysis are presented 
in Sec. III. In Sec. IV we draw our conclusions.

\section{PERTURBATIVE EXPANSIONS IN TWO DIMENSIONS}
\label{sec2}

\subsection{Renormalization of the theory}

The fixed-dimension field-theoretical approach~\cite{Parisi-80} 
represents an effective procedure in the study of the critical 
properties of systems that undergo second-order phase transitions
~\cite{ZJ-book}. In the case under consideration, the expansion is 
performed in two quartic coupling constants of the Hamiltonian 
(\ref{LGWH}). The theory is renormalized by a set of zero-momentum 
conditions for the one-particle irreducible two-point correlation 
function, four-point correlation function, and two-point correlation 
function with an insertion of the operator $\case{1}{2}\phi^2$:

\begin{equation}
\Gamma^{(2)}_{ai,bj}(p) = 
\delta_{a,b}\delta_{i,j} Z_\phi^{-1} \left[ m^2+p^2+O(p^4)\right],
\label{ren1}  
\end{equation}
\begin{equation}
\Gamma^{(4)}_{abcd}(0) = Z_\phi^{-2} m \left[  
{u\over 3}\left(\delta_{ab}\delta_{cd} + \delta_{ac}\delta_{bd} + 
  \delta_{ad}\delta_{bc} \right)
+ v \, C_{ai,bj,ck,dl}\right],
\label{ren2} 
\end{equation}
\begin{equation}
\Gamma^{(1,2)}_{ab}(0) = \delta_{ab} Z_t^{-1},
\label{ren3} 
\end{equation}
where $C_{ai,bj,ck,dl}$ is an appropriate combinatorial 
factor \cite{pp-01,prv-01}.

The perturbative knowledge of the functions $\Gamma^{(2)}$, $\Gamma^{(4)}$, 
and $\Gamma^{(1,2)}$ allows one to relate the renormalized physical parameters ($u,v,m$) to 
the bare ones ($u_0,v_0,r$ ).

The fixed points of the model are defined by the common zeros of the $\beta$ 
functions, 
\begin{eqnarray} 
\beta_u(u,v) = m \left. {\partial u\over \partial m}\right|_{u_0,v_0} ,
\quad \beta_v(u,v) = m \left. {\partial v\over \partial m}\right|_{u_0,v_0}. 
\end{eqnarray}
The stability properties of these points are determined by the eigenvalues 
$\omega_i$ of the matrix:
\begin{equation} 
\label{omega}
\Omega = \left(\matrix{\displaystyle \frac{\partial \beta_u(u,v)}{\partial u} 
&\displaystyle \frac{\partial \beta_u(u,v)}{\partial v}
 \cr \displaystyle \frac{\partial \beta_v(u,v)}{\partial u} 
& \displaystyle  \frac{\partial \beta_v(u,v)}{\partial v}}\right)\; .
\end{equation}
A fixed point is stable if both the eigenvalues have a positive real part.
If the eigenvalues possess nonvanishing imaginary parts, the fixed point 
is called a focus and the corresponding RG trajectories are spirals. The 
eigenvalues $\omega_i$ are connected to the leading scaling corrections, 
which go as $\xi^{- \omega_i}\sim |t|^{\Delta_i}=|t|^{ \nu \omega_i}$, 
with few exceptions as the two-dimensional Ising model (see for a 
discussion Ref. \cite{cccpv-00}).

The values of the critical exponents $\eta$, $\nu$ and $\gamma$ are 
related to the RG functions $\eta_{\phi}$ and $\eta_t$ evaluated at 
the stable fixed point:

\begin{eqnarray}
\eta = \eta_\phi(u^*,v^*),
\label{eta_fromtheseries} 
\end{eqnarray}
\begin{eqnarray}  
\nu = \left[ 2 - \eta_\phi(u^*,v^*) + \eta_t(u^*,v^*)\right] ^{-1},
\label{nu_fromtheseries}
\end{eqnarray}  

where 
\begin{eqnarray}
\eta_\phi(u,v) &=& \left. {\partial \ln Z_\phi \over \partial \ln m}
       \right|_{u_0,v_0}
= \beta_u {\partial \ln Z_\phi \over \partial u} + 
\beta_v {\partial \ln Z_\phi \over \partial v} ,
\end{eqnarray}
\begin{eqnarray}
\eta_t(u,v) &=& \left. {\partial \ln Z_t \over \partial \ln m}
         \right|_{u_0,v_0}
= \beta_u {\partial \ln Z_t \over \partial u} + 
\beta_v {\partial \ln Z_t \over \partial v}.
\end{eqnarray}

\subsection{Five-loop series}
 
In this section we extend the four-loop perturbative series~\cite{pp-01}
for the RG functions $\beta_u$, $\beta_v$, $\eta_\phi$, and $\eta_t$ up to 
five-loop order  using the numerical values of two-dimensional 
five-loop integrals evaluated in Ref. \cite{os-00}.

In order to obtain finite fixed point values in the limit 
of infinite components of the order parameter ($N\rightarrow \infty$) 
we use the rescaled couplings
\begin{equation}
u \equiv  {8 \pi\over 3} \; R_{2N} \; \ub,\qquad\qquad
v \equiv   {8 \pi\over 3} \;R_{2N} \; \vb ,
\label{resc}
\end{equation} 
where $R_N = 9/(8+N)$.

The resulting RG series are~\cite{foot2}: 
\begin{eqnarray} 
\bar{\beta}_{\ub} =&& -\ub + \ub^2 + {1-N\over (4+N)} \bar{u}
 \bar{v} - {1-N\over (8+2N)} \vb^2 +  \sum_{i+j\geq 3} b^{(u)}_{ij}
 \ub^i \vb^j \label{bu}, \\
\bar{ \beta}_{\vb} =&& -\vb - {1\over 2} {6-N\over (4+N)}\vb^2 +
 {6\over (4+N)} \bar{u} \bar{v} + \sum_{i+j\geq 3} b^{(v)}_{ij}
 \ub^i \vb^j 
\label{bv},  \\ 
\eta_{\phi} =&& {1.83417\,(1+N)\over (8+2N)^2} \ub^2 
+ {1.83417\,(1-N)\over (8+2N)^2} \bar{u}\bar{v} - {1.37563\,(1-N)\over 
(8+2N)^2} \vb^2+ \sum_{i+j\geq 3} e^{(\phi)}_{ij} \ub^i \vb^j
\label{etaphi},  \\
\eta_{t} =&& -{2\,(1+N)\over (4+N)} \ub  -{(1-N)\over (4+N)}\vb 
\nonumber + {13.5025\,(1+N)\over (8+2N)^2} \ub^2 + {13.5025\,(1-N)
\over (8+2N)^2} \bar{u}\bar{v} \\
&& -{10.1269\,(1-N)\over (8+2N)^2}\vb^2  
+\sum_{i+j\geq 3} e^{(t)}_{ij} \ub^i \vb^j
\label{etat},
\end{eqnarray}
where \begin{equation}
\bar{\beta}_{\ub}= {3\over 16\pi} \, R_{2N}^{-1}\beta_{u},\qquad\qquad
 \bar{\beta}_{\vb}= {3\over 16\pi} \, R_{2N}^{-1}\beta_{v}.
\label{resc2}
\end{equation}
Only  the five-loop coefficients $b^{(u)}_{ij}$, $b^{(v)}_{ij}$, $e^{(\phi)}_{ij}$ and 
$e^{(t)}_{ij}$ are reported in Table \ref{t3}, since up to the fourth order they are reported in Ref.\cite{pp-01}. 
Note that, due to the last rescaling, the element of the $\Omega$ matrix 
(\ref{omega}) are two times the derivative of $\bar{\beta}$ with respect 
to $\ub$ and $\vb$.

We verify the exactness of our series by checking them in particular limits 
($N=1$, $\bar{v}=0$, $N=M=2$, $N\rightarrow \infty$) as in Ref.~\cite{pp-01}.

\begin{table}[tbp]
\squeezetable
\caption{
Five-loop coefficients $b^{(u)}_{ij}$, $b^{(v)}_{ij}$, $e^{(\phi)}_{ij}$ and 
$e^{(t)}_{ij}$
of RG functions.}
\renewcommand\arraystretch{0.3}
\begin{tabular}{c|l}
\multicolumn{1}{c}{$i,j$}&
\multicolumn{1}{c}{$R_{2N}^{-i-j} b^{(u)}_{ij}$}\\
\tableline \hline
6,0 & $+2.01674 + 2.10643\,N+ 0.716897\,{N^2}+ 0.0887555\,{N^3}$ 
$+ 0.00241689\,{N^4} - 2.45381*10^{-7}\,{N^5} $ \\
5,1 & $ +2.92727 - 1.33251\,N - 1.34924\,{N^2} - 0.238028\,{N^3}$ 
 $ - 0.00748703\,{N^4}-0.000013119\,{N^5}$\\
4,2 & $  -4.43837 + 2.03796\,N + 2.0376\,{N^2}+0.352602\,{N^3}$ 
 $+ 0.0101649\,{N^4}+0.0000459364\,{N^5}$\\
3,3 & $+3.19454 - 1.49238\,N - 1.45358\,{N^2}- 0.242665\,{N^3}$ 
 $ - 0.0058521\,{N^4}-0.0000594059\,{N^5}$ \\
2,4 &$-1.15807 + 0.561265\,N+ 0.515855\,{N^2}+ 0.0797969\,{N^3}$ 
$ + 0.00112123 \,{N^4}+0.0000298703\,{N^5}$ \\
1,5 &$+0.222479 - 0.113395\,N - 0.095816\,{N^2} - 0.0132978\,{N^3}$ 
$+0.0000283847\,{N^4}+ 1.2973*10^{-6}\,{N^5}$ \\
0,6 &$-0.0151778 + 0.00751941\,N+ 0.00660415\,{N^2}+ 0.00104974\,{N^3}$
 $ +8.90815*10^{-6}\,{N^4}- 4.38647*10^{-6}\,{N^5}$ \\
\end{tabular}
\begin{tabular}{c|l}
\multicolumn{1}{c}{$i,j$}&
\multicolumn{1}{c}{$R_{2N}^{-i-j} b^{(v)}_{ij}$}\\
\tableline \hline
5,1 &$+6.24588+ 3.59449\,N + 0.620789\,{N^2} + 0.0264763 \,{N^3}$ 
$-0.000525164\,{N^4}-0.0000277103\,{N^5} $ \\
4,2 &$-8.7215 - 4.66521\,N - 0.686267\,{N^2}- 0.0125189\,{N^3}$ 
 $+ 0.00131132\,{N^4}+0.0000937132\,{N^5} $ \\
3,3 & $+6.32764 + 3.38603\,N + 0.514699\,{N^2}+ 0.0164667\,{N^3}$ 
$- 0.00034289\,{N^4}-0.000120039\,{N^5}$\\
2,4 & $-2.44897 - 1.35035\,N - 0.227251\,{N^2} - 0.0119867\,{N^3}$ 
 $-0.0000336496\,{N^4}+ 0.0000731821\,{N^5}$\\
1,5 &$ +0.48559 + 0.257994\,N+ 0.0396213\,{N^2}+ 0.000895835\,{N^3}$ 
$-0.000205336\,{N^4}- 0.0000218222\,{N^5}$ \\
0,6 &$ -0.036149 - 0.0191764\,N- 0.00232213\,{N^2}+ 0.000377994\,{N^3}$ 
$+0.0000915813 \,{N^4}+ 2.59344*10^{-6}\,{N^5}$ \\
\end{tabular}
\begin{tabular}{c|l}
\multicolumn{1}{c}{$i,j$}&
\multicolumn{1}{c}{$R_{2N}^{-i-j} e^{(\phi)}_{ij}$}\\
\tableline \hline
5,0 & $-0.00722954 - 0.011019\,N - 0.00391279\,{N^2}- 0.000142581\,{N^3}$ 
$-0.0000192165\,{N^4}$ \\
4,1 & $-0.0180738 + 0.00860027\,N+ 0.00916515\,{N^2}+ 0.00026037\,{N^3}$ 
$+0.0000480413\,{N^4}$ \\
3,2 & $+0.0260483 - 0.0129479\,N- 0.0129842\,{N^2}- 0.0000681925\,{N^3}$ 
$-0.0000480413\,{N^4}$ \\
2,3 &$-0.0161648 + 0.00849874\,N + 0.00785318\,{N^2} - 0.000211121\,{N^3}$
$+0.0000240207\,{N^4}$ \\
1,4 &$+0.00411472 - 0.00235806\,N - 0.00191523\,{N^2}+ 0.000164576\,{N^3}$ 
$-6.00516*10^{-6}\,{N^4}$ \\
0,5 &$-0.000397627 + 0.000247705\,N+ 0.000181639\,{N^2}- 0.0000311162\,{N^3}
- 6.00516*10^{-7}\,{N^4}$ \\
\end{tabular}
\begin{tabular}{c|l}
\multicolumn{1}{c}{$i,j$}&
\multicolumn{1}{c}{$R_{2N}^{-i-j} e^{(t)}_{ij}$}\\
\tableline \hline 
5,0 &$-0.285954 - 0.463731\,N  - 0.195487\,{N^2}  - 0.0176704\,{N^3}$ 
$+ 0.000039754\,{N^4}$ \\
4,1 &$ -0.714885 + 0.270442\,N + 0.400167\,{N^2} + 0.0443747\,{N^3}$
$- 0.000099385\,{N^4}$ \\
3,2 &$+0.954626 - 0.384666\,N - 0.520913\,{N^2}  - 0.0491461\,{N^3}$ 
$+ 0.0000993852\,{N^4}$ \\
2,3 &$-0.572386 + 0.253092\,N + 0.29846\,{N^2}  + 0.0208835\,{N^3}$ 
$- 0.0000496928\,{N^4}$ \\
1,4 &$+0.149502 - 0.0775891\,N- 0.0701865\,{N^2}- 0.00179722\,{N^3}$ 
 $+0.0000708398\,{N^4}$  \\
0,5 &$-0.0157319 + 0.0101789\,N + 0.00583765\,{N^2}  - 0.00022754\,{N^3}$ 
$-0.0000571742\,{N^4}$  \\
\end{tabular}
\label{t3}
\end{table}

\section{RESUMMATION AND ANALYSIS OF THE FIVE-LOOP SERIES}
\label{sec3}

The field theoretical perturbative expansions are known to be divergent 
asymptotic series. Quantitative informations may be extracted from these 
series exploiting the properties of Borel summability for $\phi^4$ 
theories in two and three dimensions and resumming them by a Borel 
transformation combined with a method for the analytical extension of 
the Borel transform. This last procedure can be realized by using 
Pad\'e approximants or via a conformal mapping, which maps the domain 
of analyticity of the Borel transform (cut at the instanton singularity) 
$\ub_b$ onto a circle \cite{L-Z-77}.

Let us consider a perturbative series in $\ub$ and $\vb$ 
\begin{equation}
R(\ub,\vb)= \, \sum_{k=0}^l \sum_{h=0}^{l-k} R_{hk} \ub^h \vb^k,
\end{equation}
which we want to resum (i.e., one of the $\eta$ or $\beta$ functions).
In the Pad\'e-Borel method we use the trick of the resolvent series 
considering the Borel transformed function, 
\begin{equation} 
\tilde{R}(\lambda, \ub,\vb,b)=\, \sum_{n=0}^{\infty}{\lambda^n \over 
\Gamma(n+b+1)} \sum_{l=0}^n R_{l,n-l} \ub^l \vb^{n-l}. 
\end{equation} 
Calling $P^{N}_M(\lambda,\ub,\vb,b)$ the $[N/M]$ Pad\'e approximant 
of the series $\tilde{R}(\lambda, \ub,\vb,b)$ in the variable 
$\lambda$, one obtains the estimate of the desired quantity as 
\begin{equation} 
\tilde{P}^{N}_M(b,\ub,\vb)=\, \int_0^\infty e^{-t}t^b P^{N}_M(t,\ub,\vb,b) dt.
\end{equation}
In this way we produce several approximants of the function $R(\ub,\vb)$ 
with varying the parameters $b$, $N$, and $M$.

In the conformal mapping method, the idea is to exploit, in the course of
the resummation, the knowledge of the large order behavior of the 
series $R(\ub,\vb)$ at fixed $z=\vb/\ub$~\cite{cpv-00,prv-01,pp-01}.
This large-order behavior is related to the singularity of 
the Borel transform closest to the origin~\cite{B-L-Z-77}.
 For the Hamiltonian (\ref{LGWH}), the value of the Borel transform 
singularity closest to the origin $\ub_b$ has been computed in 
Ref~\cite{pp-01}, obtaining:

\begin{eqnarray}
{1\over \ub_b} =& - a\, R_{2N}
\qquad & {\rm for} \qquad  0< z<4 ,
\label{bsing} \\
{1\over \ub_b} =& - a\, R_{2N} \,\left( 1  - \displaystyle 
{1\over 2} z \right) \qquad & {\rm for} \qquad   z<0 ,\quad z>4,
\nonumber
\end{eqnarray}
where $a = 0.238659217\dots$. For $z>2$ there is a singularity on the 
real positive axis which, however, is not the closest one to the origin 
for $z<4$. Thus, for $z>2$ the series are not Borel summable.
We resum the RG functions also for the values of $z$ for which the series 
are not Borel summable since the method should provide a reasonable estimate
if $z<4$, because we take into account the leading large-order behavior.

With the knowledge of $\ub_b$, one can perform the mapping 
\begin{equation} 
y(\ub;z) = {\sqrt{1 - \ub/\overline{u}_b(z)} - 1\over 
          \sqrt{1 - \ub/\overline{u}_b(z)} + 1}, 
\end{equation} 
in order to extend $R(\ub,\ub z)$ to all positive values of $\ub$. 
Consequently one obtains the set of approximants
\begin{equation}
E({R})_p(\alpha,b;\ub,\vb)= \sum_{k=0}^p 
  B_k(\alpha,b;\vb/ \ub) \times   \int_0^\infty dt\,t^b e^{-t} 
  {y(\ub t;\vb/\ub)^k\over [1 - y(\ub t;\vb/ \ub)]^\alpha},
\label{approx}
\end{equation}
where the coefficient $B_k$ is determined by the condition that the 
expansion of $E({R})_p(\alpha,b;\ub,\vb)$ in powers of $\ub$ and $\vb$ 
gives $R(\ub,\vb)$ to order $p$.

An important issue in the fixed dimension approach to critical phenomena 
concerns the analytic properties of the $\beta$ functions. 
As shown in Ref. \cite{cccpv-00} for the $O(N)$ model, the presence of 
confluent singularities in the zero of the perturbative $\beta$ function 
causes a slow convergence of the resummation of the perturbative series 
to the correct fixed-point value. 
The apparent stability of the results when analyzing a finite number of terms 
of the perturbative expansion may not provide a reliable indication of the 
uncertainty of the overall estimates. 
The $O(N)$ two-dimensional field-theory 
valuations of physical quantities \cite{L-Z-77,os-00} are less accurate 
than the three-dimensional counterparts, due to the bigger critical value 
of the quartic coupling constant and stronger nonanalyticities at the
 fixed point \cite{cccpv-00,nickel-82,pv-98}. 
In Ref. \cite{cccpv-00} it is explicitly 
shown that the nonanalytic terms cause a large deviation in the estimate 
of the right correction to the scaling, i.e., the exponent $\omega$. At 
the same time, the perturbative results for the fixed-point location turn 
out to be rather good approximations for the exact ones. We think that a 
similar situation holds for frustrated models.

\subsection{Fixed points, stability, and critical exponents}
\label{fpoints}

\begin{figure}[t]
\centerline{\psfig{height=4truecm,width=8.6truecm,file=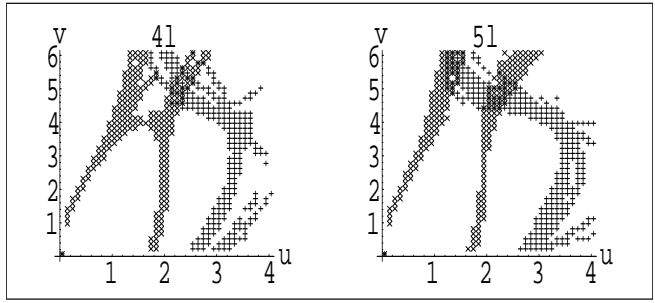}}

\caption{Zeros of $\bar{\beta}$ functions for $N=2$ in the $(\ub,\vb)$ 
plane. Pluses ($+$) and crosses ($\times$) correspond to zeroes of 
$\bar{\beta}_{\vb}(\ub,\vb)$ and $\bar{\beta}_{\ub}(\ub,\vb)$, respectively.}
\label{n2}
\end{figure}

\begin{figure}[t]
\centerline{\psfig{height=4truecm,width=8.6truecm,file=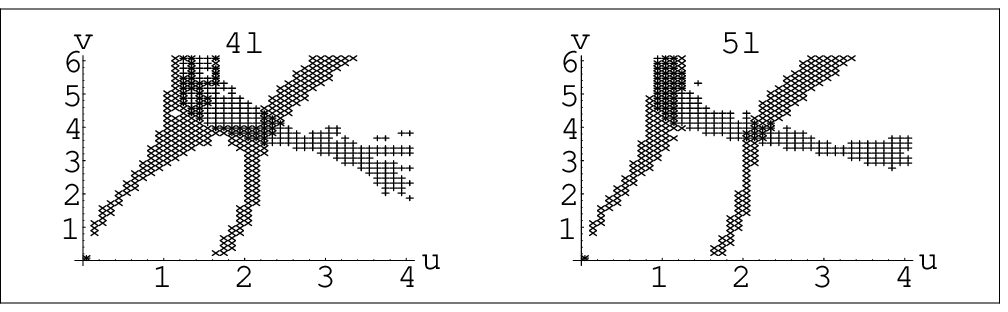}}

\caption{Zeros of $\bar{\beta}$ functions for $N=3$ in the $(\ub,\vb)$ 
plane. Pluses ($+$) and crosses ($\times$) correspond to zeroes of 
$\beta_{\vb}(\ub,\vb)$ and $\bar{\beta}_{\ub}(\ub,\vb)$, respectively.}
\label{n3}
\end{figure}

In order to calculate the fixed points of the theory, we resum the 
perturbative expression for each $\beta$ function in the whole plane 
$(\ub,\vb)$ following the procedure explained above.
 With the conformal mapping method, we choose the approximants 
that stabilize the series for small values of the coupling constants 
with varying the considered perturbative order, i.e., $\alpha=0,1,2$ and 
$b=5,7,\dots ,15$. In this manner we have $18$ approximants for each $\bar{\beta}$ function. 
To obtain reasonable values and error bars for the fixed-point coordinates, 
we divide the domain $0\leq \ub\leq 4$, $0\leq \vb\leq 6$ in $40^2$ 
rectangles, and we mark all the sites in which at least two approximants 
for $\bar{\beta}_{\ub}$ and $\bar{\beta}_{\vb}$ vanish. This procedure is 
applied for the four-loop and five-loop series and the results for the 
zeros of the $\beta$ functions for $N=2$ and $N=3$ are displayed in Figs.~\ref{n2} and \ref{n3}, respectively.
We remind the reader that the four-loop results were previously reported in 
Ref. \cite{pp-01}, where similar figures were shown also for the 
three-loop case. It was also shown that no upper branch of zeros of 
$\bar{\beta}_\ub$ appears in the three-loop approximation for all 
considered values of $N$.

As one can see from Figs.~\ref{n2} and \ref{n3}, for these physical systems 
the resummed RG expansions yield similar pictures for the zeros of 
the $\bar{\beta}$ functions. The existence of three fixed points -- the 
gaussian, the $O(2N)$, and the chiral -- is clear in both cases, while 
the presence of an unstable antichiral fixed point looks doubtful. 
In fact, if this last point exists, it should be located in the domain 
where the resummation procedure fails to give reliable quantitative 
results ($z>4$). 

\begin{table}[b]
\caption{
Location of the chiral fixed point with its stability eigenvalues, 
critical exponent $\eta$, and crossover exponent $\omega_{O(2N)}$ at 
the $O(2N)$ fixed point.
}
\begin{tabular}{l|l|l|l}
$N=2$ & Pad\'e & C.M. (4 loop) & C.M. (5 loop)\\
\tableline \hline
$\ub^*$  &   2.35(30)  & 2.3(2) & 2.25(25) \\ 
$\vb^*$  &   4.35(25) & 5.0(5) & 4.80(45) \\
$\omega_{\pm}$   & $2.33(65) \pm i0.62(48)$ & $1.50(25) \pm i 1.00(45)$ 
& $2.05(35)\pm i 0.80(55)$ \\
$\eta$       &   0.3(1)         &  0.29(5) & 0.28(8)\\
$\omega_{O(4)}$ &           &-0.36(4) & -0.35(16)      \\
\tableline \hline
 $N=3$ & Pad\'e & C.M.(4 loop) & C.M.(5 loop)\\
\tableline \hline 
$\ub^*$  &2.25(20)     & 2.3(3) & 2.2(2) \\ 
$\vb^*$  & 3.60(15)   & 3.9(5) & 3.9(3) \\
$\omega_{\pm}$ &$1.77(43)\pm i0.47(18)$ & $1.30(25) \pm i0.50(35)$ 
& $1.55(25)\pm i 0.55(35)$ \\
$\eta$       &   0.23(7)         &  0.24(6) & 0.23(5) \\
$\omega_{O(6)}$ &           &   -0.7(2)       & -0.69(9) \\
\end{tabular}
\label{tabMC2}
\end{table}

The location of the chiral fixed point for $N=2$ is 
\begin{eqnarray} 
[u^*,v^*]=& [2.3(2), 5.0(5)]      & \qquad(\mbox{4-loop CM}),\\
          & [2.25(25), 4.80(45)]  & \qquad(\mbox{5-loop CM}),\\
          & [2.35(30), 4.35(25)]  & \qquad(\mbox{Pad\'e}),
\end{eqnarray}
and for $N=3$ 
\begin{eqnarray}
[u^*,v^*]=& [2.3(3), 3.9(5)]      & \qquad(\mbox{4-loop CM}),\\
          & [2.2(2), 3.9(3)]      & \qquad(\mbox{5-loop CM}),\\
          & [2.25(20), 3.60(15)]  & \qquad(\mbox{Pad\'e}).
\end{eqnarray}

The stability properties of the chiral fixed point depend on the signs 
of the real parts of the eigenvalues $\omega_i$ of the $\Omega$ matrix. 
We take the above 18 
approximants  for each $\bar{\beta}$ function 
 and compute the numerical derivatives of each couple of 
approximants of the two $\bar{\beta}$ functions at their common zero, 
obtaining 
324 possible combinations. To have reliable numerical estimates for 
$\omega_i$, we limit ourselves by the approximants that yield the chiral 
fixed point coordinates compatible with their final, properly weighted 
values.

Both for $N=2$ and $N=3$, we find that the chiral fixed point is a 
stable focus since it possesses stability eigenvalues with nonvanishing 
imaginary parts and positive real parts. At the five-loop (four-loop) 
approximations for $N=2$ the complex eigenvalues are produced by 87\% 
(87\%) of the working approximants within the conformal mapping method, 
while for $N=3$ by 89\% (74\%) of the approximants employed. 
When estimating the averaged eigenvalues of the $\Omega$ matrix at the 
chiral fixed point, we disregard the approximants leading to purely 
real eigenvalues. For $N=2$ the 
final results of our calculations are: 
\begin{eqnarray} 
\omega_{\pm}=& 1.50(25)\pm i \,1.00(45), & \quad \mbox{(4-loop CM)}  \\
\omega_{\pm}=& 2.05(35)\pm i \,0.80(55), & \quad \mbox{(5-loop CM)}  
\end{eqnarray}
and for $N=3$
\begin{eqnarray}
\omega_{\pm}=& 1.30(25)\pm i \,0.50(35),  & \quad \mbox{(4-loop CM)}  \\
\omega_{\pm}=& 1.55(25)\pm i \,0.55(35),  & \quad \mbox{(5-loop CM)}
\end{eqnarray}

The presence of imaginary parts comparable in magnitude with the real 
ones leads to the conclusion that the critical behaviors of frustrated 
two-dimensional Heisenberg and $XY$ systems are driven by focus fixed 
points. To strengthen this new and important issue, 
we repeat the analysis of the stability properties 
of the chiral fixed point by a Pad\'e study, which confirms the previous 
results, although the statistics in this case is much less significant 
due to a large number of defective Pad\'e approximants. All the results 
obtained are summarized in Table \ref{tabMC2}.
Note that in the past such a behavior was found in three dimensions within 
the three-loop RG approximation but only for unphysical values of $N$ 
($N = 5, 6, 7$)~\cite{lsd-00}.

In order to fully characterize the two physical systems discussed, we 
calculate the critical exponents as well. For the exponent $\eta$ we 
find values that agree with the results of the four-loop analysis
~\cite{pp-01}. At the same time, we can not give a reasonable estimate 
for the critical exponent $\nu$ because of the strong oscillations 
observed with varying the working approximants.

Finally, we report in  Table \ref{tabMC2} the 
five-loop estimate of the crossover exponent $\omega_{O(2N)}$, 
finding that the Heisenberg fixed point is unstable for both $N=2,3$,
in agreement with the previous four-loop calculation \cite{pp-01}.

We apply the same kind of analysis also to the models with $N\geq 4$.
All these systems  are free of topological defects~\cite{pp-01}
and are well described by appropriate $NL\sigma$ models \cite{nlsigma}
which provide the existence of a stable chiral fixed point with critical 
exponents and corrections to the scaling equal to the ones of a system 
with an infinite number of components of the order parameter.

\begin{figure}[tcb]
\centerline{\psfig{height=4truecm,width=8.6truecm,file=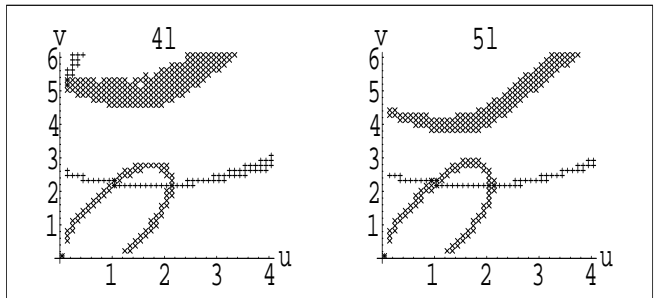}} 
\caption{Zeros of $\bar{\beta}$ functions for $N=32$ in the $(\ub,\vb)$ 
plane. Pluses ($+$) and crosses ($\times$) correspond to zeroes of 
$\bar{\beta}_{\vb}(\ub,\vb)$ and $\bar{\beta}_{\ub}(\ub,\vb)$, respectively.}
\label{n32}
\end{figure}

\begin{figure}[tcb]
\centerline{\psfig{height=4truecm,width=8.6truecm,file=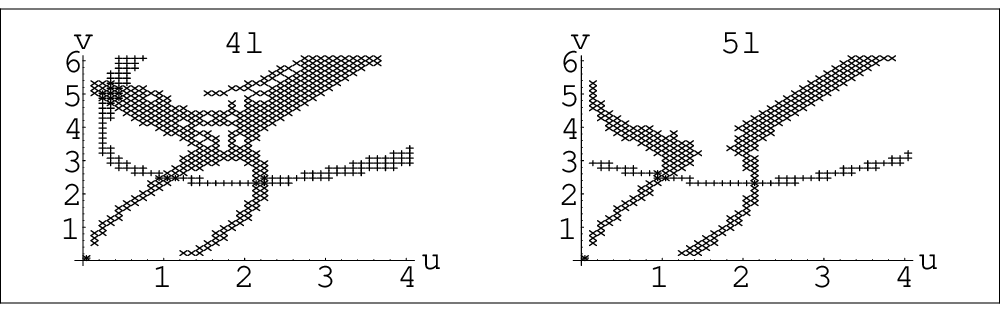}}
\caption{Zeros of $\bar{\beta}$ functions for $N=16$ in the $(\ub,\vb)$ plane.
Pluses ($+$) and crosses ($\times$) correspond to zeroes of 
$\bar{\beta}_{\vb}(\ub,\vb)$ and $\bar{\beta}_{\ub}(\ub,\vb)$, respectively.}
\label{n16}
\end{figure}

\begin{figure}[tcb]
\centerline{\psfig{height=4truecm,width=8.6truecm,file=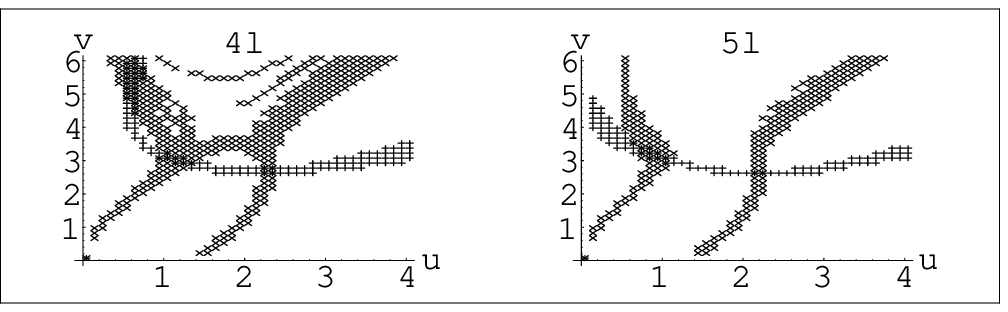}}
\caption{Zeros of $\bar{\beta}$ functions for $N=8$ in the $(\ub,\vb)$ plane.
Pluses ($+$) and crosses ($\times$) correspond to zeroes of 
$\bar{\beta}_{\vb}(\ub,\vb)$ and $\bar{\beta}_{\ub}(\ub,\vb)$ respectively.}
\label{n8}
\end{figure}

\begin{figure}[tcb]
\centerline{\psfig{height=4truecm,width=8.6truecm,file=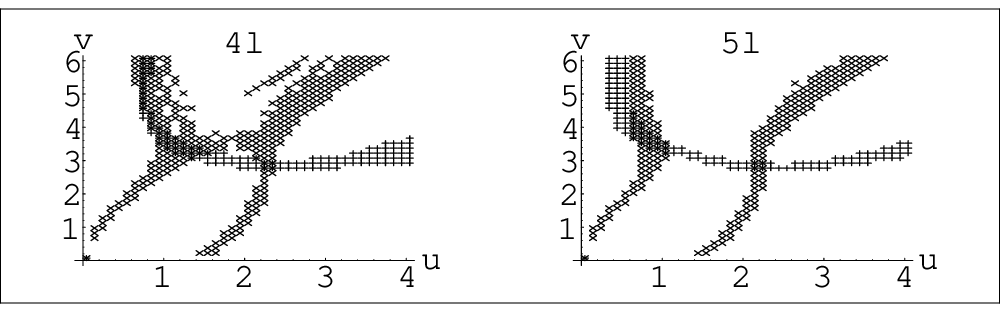}}
\caption{Zeros of $\bar{\beta}$ functions for $N=6$ in the $(\ub,\vb)$ plane.
Pluses ($+$) and crosses ($\times$) correspond to zeroes of 
$\bar{\beta}_{\vb}(\ub,\vb)$ and $\bar{\beta}_{\ub}(\ub,\vb)$ respectively.}
\label{n6}
\end{figure}

\begin{figure}[tcb]
\centerline{\psfig{height=4truecm,width=8.6truecm,file=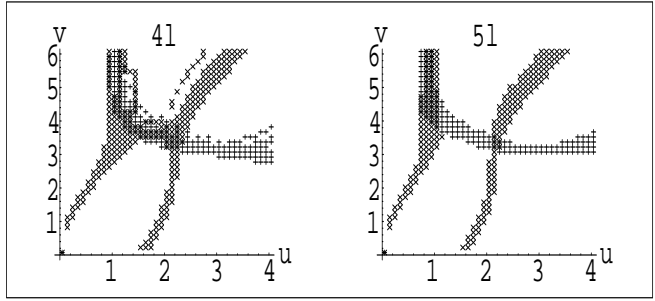}}
\caption{Zeros of $\bar{\beta}$ functions for $N=4$ in the $(\ub,\vb)$ plane.
Pluses ($+$) and crosses ($\times$) correspond to zeroes of 
$\bar{\beta}_{\vb}(\ub,\vb)$ and $\bar{\beta}_{\ub}(\ub,\vb)$ respectively.}
\label{n4}
\end{figure}

The zeros of the two $\beta$s for 
$N=32, 16, 8, 6, 4$ are shown in 
Fig.  \ref{n32},  \ref{n16}, \ref{n8}, \ref{n6}, \ref{n4}.
From these figures it is clear that for large enough values of $N$ 
four fixed points exist: the gaussian, the $O(2N)$, the chiral, and the 
antichiral, with a topological structure similar to the one found in 
$\epsilon$ expansion for $N>N_c$. The location of the chiral fixed point 
for various $N$ is reported in Tab. \ref{tabellone}.

From the $NL\sigma$ model it is expected $\omega_1= \omega_2=2$ \cite{nlsigma}.
For this reason $N^*$ should be less or at least equal to $4$.
From our analysis we find that the chiral fixed point is stable for all the 
considered $N$. However, the 
estimates for $\omega_i$ are slighty different from $2$, 
also for large $N$~(e.g., for $N=16$ we find 
$\omega_1=1.59(5)$ and $\omega_2=2.01(11)$). 
Such a disagreement was 
already observed in the four-loop approximation, and it may be attributed 
to the nonanalyticities discussed in the previous subsection. 
We observe that for $N\geq 8$ all the stability eigenvalues are 
purely real, but for $N= 6$ about the $50\%$ of the five-loop approximants 
possess a nonvanishing imaginary part, suggesting $N^*\sim6$. 
This result seems to disagree with the $NL\sigma$ prediction. 
Again this might be due  to nonanalyticities.

\begin{table}[t]
\caption{Location of the stable chiral fixed point for $N\geq 4$.}
\begin{tabular}{c|ccc}
$N$ & C.M.(4 loop) & C.M.(5 loop) & Pad\'e  \\
\hline
4   &                    & [2.2(1),3.37(23)] & [2.22(16),3.19(10)]\\
6   & [2.26(7),2.85(15)] & [2.21(6),2.80(8)] &[2.25(10),2.79(4)] \\
8   & [2.28(9),2.66(8)]  &[2.22(7),2.61(6)]  &[2.23(6),2.61(3)]  \\
16  & [2.22(4),2.37(3)]  &[2.17(5),2.33(3)]  &[2.18(2),2.33(3)]  \\
32  & [2.128(16),2.19(1)]&[2.09(2),2.16(2)]  &[2.100(15),2.165(10)] \\
\end{tabular}
\label{tabellone}
\end{table}

To further approve the above conclusions, we apply, along with the 
conformal-mapping based, the resummation by means of the Pad\'e-Borel 
technique. In the region where the fixed point is expected, many Pad\'e 
approximants turn out to be defective, i.e., possessing dangerous poles. 
We find that the approximants $[4/1]$, $[3/1]$, and $[2/2]$ are 
nondefective in the majority of the cases. Considering 
these three Pad\'e approximants 
with $b=0,1,2$ (being the more stable 
under variation of the number of loops), we finally have nine approximants 
for each $\beta$ function. The location of the chiral fixed point is 
reported in Table \ref{tabellone}, with all the defective cases discarded.

We also compute 
the exponents $\eta$ and $\eta_t$ with both the resummation techniques. 
We consider a lot of approximants and finally we choose the more stable 
ones with varying the number of loops. 
The found values are close to those of the 
$O(N)$ model for $N\geq 3$, but differ from the exact ones coming 
from $NL \sigma$ model $\eta=0$ and $\eta_t=-2$ (e.g., we find $\eta_t=-1.95(20)$
for $N=16$ and $\eta=0.16(2)$ for $N=6$).
We think  these discrepancies are due to 
nonanalyticities, as in the case of $O(N)$ models.
To conclude the analysis of the fixed points, we study the stability 
of the Heisenberg fixed point by calculating the crossover exponent 
$\omega_{O(2N)}$. The five-loop results 
show that this fixed point is unstable, 
in agreement with the conclusion drawn earlier from the four-loop 
series~\cite{pp-01}.

\subsection{Renormalization-group flow and crossover}

\begin{figure}[b]
\centerline{\epsfig{height=6truecm,width=8.6truecm,file=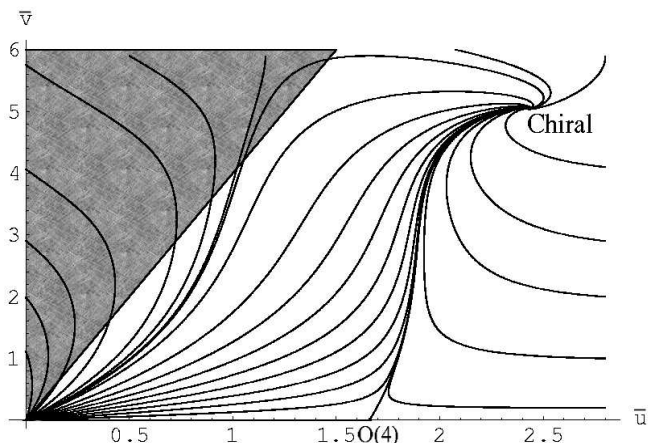}}
\caption{RG flow for the two-dimensional chiral model with $N=2$. The 
approximants used are given by $\alpha=1$ and $b=5$ for $\beta_\ub$ and 
$\alpha=0$ and $b=8$ for $\beta_\vb$. In this case, the chiral fixed 
point is at $(2.427,5.045)$.}
\label{fign2d2}
\end{figure}

\begin{figure}[t]
\centerline{\psfig{height=4truecm,width=8.6truecm,file=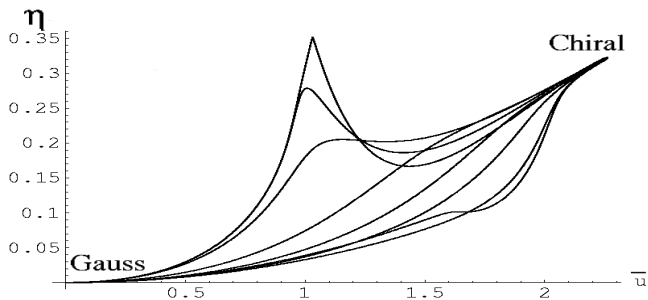}}
\caption{Crossover of the effective exponent $\eta_{eff}$ for $N=2$.}
\label{cr}
\end{figure}

In this subsection we demonstrate the structure of the RG flow for 
physical values of $N$. As it was already shown, a stable focus governs 
the critical behavior of these systems. In Fig. \ref{fign2d2} ($N=2$)
one can see the spiral-like approach of the stable fixed point which 
is peculiar for the fixed points with stability eigenvalues possessing 
nonvanishing imaginary parts. 
The shaded area is the region where the resummation procedure is expected to 
fail (i.e., $z>4$), since we do not take into account the singularity of the 
Borel transform closest to the origin that is on the real positive axis (cf. 
Eq (\ref{bsing})).
Consequently, the RG trajectories given by the resummed perturbative series are
quantitatively correct only within the unshaded areas. We report, nevertheless,
the RG flows in other parts of the coupling constant plane in order to present 
a complete qualitative picture.
Note, that the domain usually referred to as the sector of the first-order 
transitions, where the positiveness of the quartic form in the free-energy 
expansion breaks down, is given by the inequality $z>2$. In fact, however, 
because of the presence of the higher-order terms in this expansion keeping 
the system globally stable at any temperature, the true domains of the 
first-order transitions should be more narrow.

From Fig. \ref{fign2d2} we see that for some initial values of $(\ub,\vb)$ 
the RG trajectories have the coordinate $\vb$ that grows very fast at 
the beginning and seems to reach the region of the first-order phase 
transitions, but just before arriving there these trajectories drastically 
curve moving toward the stable chiral fixed point. These bizarre  
trajectories can imitate both the discontinuous phase transitions and 
strongly scattered effective critical exponents observed in numerous 
physical and computer experiments. To substantiate the former scenario,
we calculate the numerical values which the effective exponent 
$\eta_{eff}$ takes along the RG trajectories running from the Gaussian 
fixed point to the chiral one. Corresponding curves are shown 
in Fig. \ref{cr}. As is seen from this figure, the exponent oscillates 
within a large range, even near the stable fixed point.

\begin{figure}[t]
\centerline{\epsfig{height=6truecm,width=8.6truecm,file=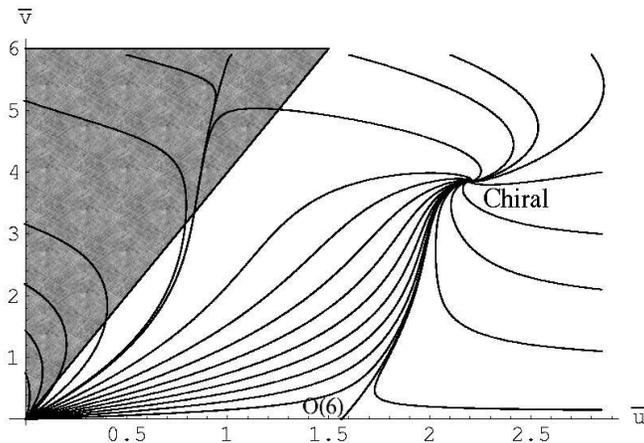}}
\caption{RG flow for the two-dimensional chiral model with $N=3$. Both  
$\beta$ functions are given by approximants given by $\alpha=2$ and $b=9$, 
the chiral fixed point is located at $(1.702, 2.858)$.}
\label{fign3d2}
\end{figure}

\begin{figure}[t]
\centerline{\epsfig{height=6truecm,width=8.6truecm,file=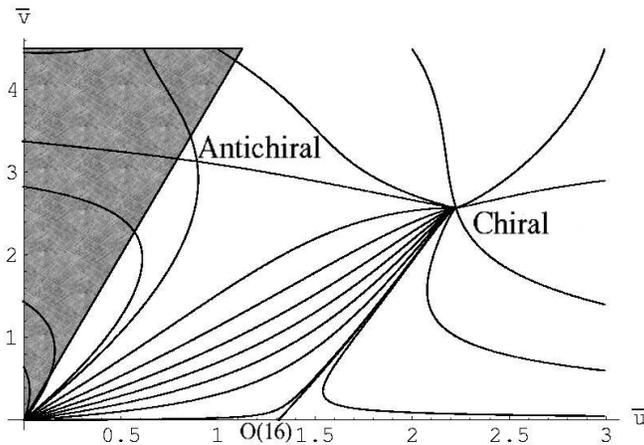}}
\caption{RG flow for the two-dimensional model with $N=8$. 
The approximants used are given by $\alpha=0$ and $b=7$ for $\beta_\ub$ 
and $\alpha=1$ and $b=5$ for $\beta_\vb$. In this case, the chiral fixed 
point is at $(2.226,2.571)$ and the antichiral fixed point is at 
$(0.891,3.133)$.}
\label{f8}
\end{figure}

\begin{figure}[t]
\centerline{\psfig{height=4truecm,width=8.6truecm,file=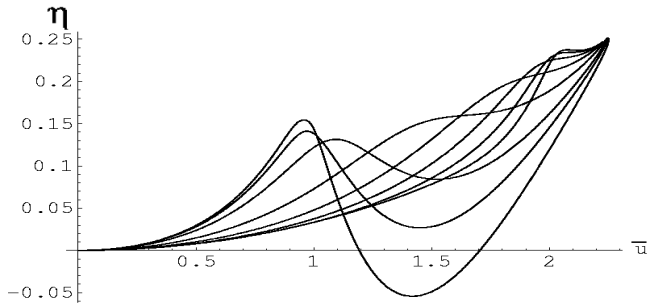}}
\caption{Crossover of the effective exponent $\eta_{eff}$ for $N=3$.}
\label{crn3}
\end{figure}

We demonstrate also the RG flows for $N=3$ (Fig. \ref{fign3d2}) 
and for $N=8>N^*$ (Fig. \ref{f8}). In the former (physical case) the 
stable fixed point is a focus and phase trajectories approach it in a 
spiral-like manner. In the latter (unphysical case) the stable fixed 
point turns out to be a node and the trajectories are certainly not 
spirals. Also for $N=3$ we estimate the effective exponent 
$\eta_{eff}$ along the RG trajectories running from the Gaussian 
fixed point to the chiral one. The corresponding curves (cf. Fig. \ref{crn3}) 
show big oscillations and sometimes reach negative values.
We note that these negative values are obtained only in the region where the
resummation is not sure, i.e., $z>4$. 
 Nevertheless, this strange crossover, if reproduced from the three-dimensional 
series \cite{CPS-02b}, might explain some doubtful experimental results, 
where a negative $\eta$ was found.

To conclude, 
it is worthy to note that earlier the focuslike stable fixed points were
found on the RG flow diagrams of the model describing critical behavior
of liquid crystals \cite{lc} and of the $O(n)$-symmetric systems undergoing
first-order phase transition close to the tricritical point \cite{tr}.
In those cases, however, the independent coupling constants had different 
scaling dimensionality and played essentially different roles in  
critical thermodynamics. 
Moreover, recently complex subleading exponents were found from
the high-temperature expansion of Dyson's hierarchical Ising 
model \cite{mor-95}.

\section{CONCLUSIONS}

We studied the critical thermodynamics of the two-dimensional $N$-vector 
chiral model in the five-loop RG approximation. Using the advanced 
resummation technique based upon Borel transformation combined with conformal 
mapping and use of Pad\'e approximants, 
it was shown that the fixed point governing the chiral critical 
behavior is a stable node only for large enough $N$, $N > N^*$. For smaller 
$N$ a chiral fixed point turns out be a stable focus and the RG 
trajectories approach this point in a spiral-like manner. Of particular 
importance is the fact that a focus-driven critical behavior was found 
for systems with physical values of $N$, $N = 2, 3$. 
Evaluating the critical exponent $\eta$ along some RG 
trajectories, we demonstrate that spiral-like approach of the chiral 
fixed point results in unusual crossover and near-critical regimes 
that may imitate varying critical exponents seen in physical and 
computer experiments.

\section*{ACKNOWLEDGMENTS}
We would like to thank Ettore Vicari for many useful discussions.  
The financial support of the Russian Foundation for Basic 
Research under Grant No. 01-02-17048 (E.V.O., A.I.S.) and the 
Ministry of Education of Russian Federation under Grant 
No. E00-3.2-132 (E.V.O., A.I.S.) is gratefully acknowledged. 
A.I.S. has benefited from the warm hospitality of Scuola 
Normale Superiore and Dipartimento di Fisica 
dell'Universit\`a di Pisa, where the major part of this 
research was done.


\end{document}